\documentclass{icrc2009}
 
\usepackage{graphicx}   
\usepackage[caption=false]{caption}    
\usepackage[font=footnotesize]{subfig} 
\usepackage{fixltx2e}
\usepackage{url}

\newcommand{\shorttitle}[1]%
{\markboth{Proceedings of the 31\MakeLowercase{$^{st}$} ICRC, {\L}\'{o}d\'{z} 2009}{#1} }
\newcommand{\etal}{\MakeLowercase{\textit{et al. }}} 

\usepackage{amsmath}
\usepackage{amssymb}

\newcommand{\bq}{\begin{equation}}
\newcommand{\eq}{\end{equation}}
\newcommand{\bqan}{\begin{eqnarray*}}
\newcommand{\eqan}{\end{eqnarray*}}

\begin{document}
\title{The starburst-GRB connection}

\author{\IEEEauthorblockN{Jens Dreyer\IEEEauthorrefmark{1}\IEEEauthorrefmark{2}, Julia K. Becker\IEEEauthorrefmark{1}\IEEEauthorrefmark{2} and Wolfgang Rhode\IEEEauthorrefmark{1}}
\IEEEauthorblockA{\IEEEauthorrefmark{1}Technische Universit\"{a}t Dortmund, Fakult\"{a}t f\"{u}r Physik, D-44221 Dortmund, Germany}
\IEEEauthorblockA{\IEEEauthorrefmark{2}Institut f\"ur Theoretische Physik, Ruhr-Universit\"at Bochum, D-44780 Bochum, Germany}}


\shorttitle{Jens Dreyer \etal Starburst-GRB connection}
\maketitle

\begin{abstract}
As starburst galaxies show a star formation rate up to several hundred times larger than the one in a typical galaxy, the expected supernova rate is higher than average. 
This in turn implies a high rate of long gamma ray bursts (GRBs), which are extreme supernova events. We present a catalog of 127 local starburst galaxies with redshifts of 
$z<0.03$. Using this catalog we investigate the possibility of detecting neutrinos from Gamma Ray Bursts from nearby starburst galaxies. We show that the rate of long GRBs is correlated to 
the supernova rate which in turn is correlated to the far infrared
output. For the entire catalog, $0.03$ GRB per year are expected to
occur. The true number can even be higher since only the brightest sources were included in the catalog. 
\end{abstract}

 
\section{Introduction}
In our work \cite{CR6} we present a systematic investigation of radiative emission from starburst galaxies. We explain there the correlation between
radio emission and emission in the far infrared and high energy photons and neutrinos. In this contribution we will focus on
the possibility to detect secondaries from gamma ray bursts in starburst galaxies. In section \ref{sample} the sample of 127 starburst galaxies used
will be presented. Section \ref{grb_sb} explains the connection between starburst galaxies and gamma ray bursts while section
\ref{nuflux} presents the neutrino spectrum for GRBs. Expected event
rates in the IceCube detector for GRBs from nearby starburst galaxies are shown in sub section \ref{ic3}.  
\section{A local starburst sample}
\label{sample}
First a sample of $127$ nearby starburst galaxies is seleected. For the data of the individual sources see \cite{CR6} Appendix A.
This is a local sample containing only sources with $z<0.03$. The sources were selected from a larger catalog of starbursts which
contained $309$ sources. To ensure  completeness of the sample, cuts in the radio flux as well as in the FIR flux were applied. To ensure that the selected
sources were indeed starbursts, only sources with a high ratio of FIR flux and radio flux, $S_{60\mu} / S_{1.4\,\rm{GHz}}>30$ were selected.
$S_{60\mu}$ denotes the FIR flux density at $60\,\mu\rm{m}$ wavelength measured by the IRAS satellite, $S_{1.4\,\rm{GHz}}$ denotes the radio
flux density at $1.4\,\rm{GHz}$. This criterium removes possible contamination of the sample by Seyfert galaxies. The ratio is $\sim 10$ for Seyfert
galaxies and $\sim 300$ for starbursts.
Further, luminosity cuts were applied. It was required that $S_{1.4\,\rm{GHz}}>20\,\mbox{mJy}$\footnote{$1$ Jansky (Jy) $= 10^{-26}\,\frac{\mbox{W}}{\mbox{Hz} \cdot \mbox{m}^2}$} 
and $S_{60\mu}>4\,\mbox{Jy}$. 
\begin{figure*}[!t]
\centerline{\subfloat[Luminosity at $60\,\mu\mbox{m}$ versus the luminosity distance]{\includegraphics[width=\columnwidth]{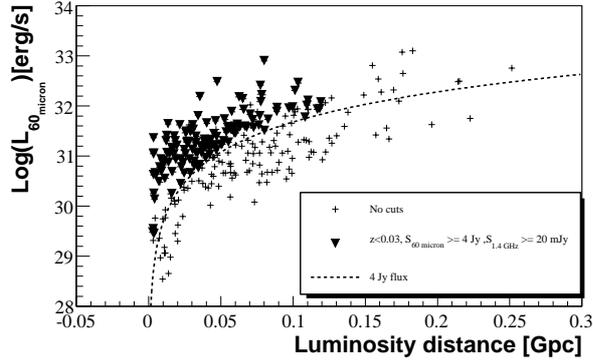} \label{LDL}}
            \hfil
            \subfloat[Luminosity at $1.4\,\mbox{GHz}$ versus the luminosity distance]{\includegraphics[width=\columnwidth]{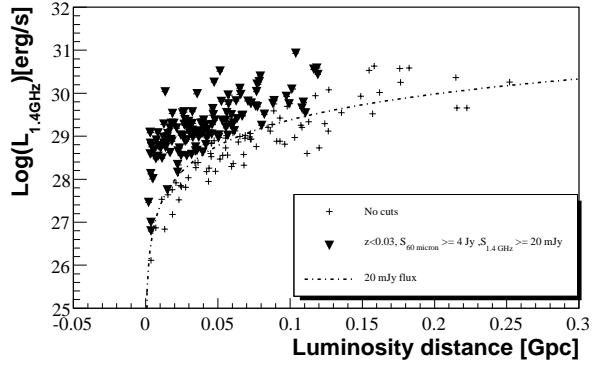} \label{LDLR}}
            }
\caption{Luminosity-distance diagrams. Crosses denote all $309$ pre-selected starburst galaxies, triangles show the remaining after the cuts
$S_{60\mu}>4\,\mbox{Jy}$ and $S_{1.4\,\rm{GHz}}>20\,\mbox{mJy}$ and $z<0.03$. The dashed lines show the sensitivity for $S_{60\mu}=4\,\mbox{Jy}$ and $S_{1.4\,\rm{GHz}}=20\,\mbox{mJy}$ respectively.}
\label{ldldiag}
\end{figure*}

Figure \ref{LDL} and figure \ref{LDLR} show the luminosity of the starburst galaxies versus their luminosity distance at $60\,\mu\mbox{m}$ resp. at $1.4\,\mbox{GHz}$.
The crosses represent all $309$ starbursts that were selected in the beginning, the triangles show the remaining $127$ sources after the cuts $S_{60\mu}>4\,\mbox{Jy}$ and $S_{1.4\,\rm{GHz}}>20\,\mbox{mJy}$ and $z<0.03$.
The dotted lines represent the sensitivity for $4\,\mbox{Jy}$ and $20\,\mbox{mJy}$, respectively with those cuts applied, a complete, local sample in both wavelengths, FIR and radio is obtained.
Since all sources in the sample have a redshift of $z<0.03$, many of them are located in the super galactic plane. Therefore, the number of sources with a flux density larger than $S$, $N(>S)$ is expected to
follow a powerlaw behavior of $S^{-1}-S^{-1.5}$. For a flat cylinder a pure $S^{-1}$ behavior is expected while a spherical distribution results in an $S^{-1.5}$ behavior.
\begin{figure}
\includegraphics[width=\columnwidth]{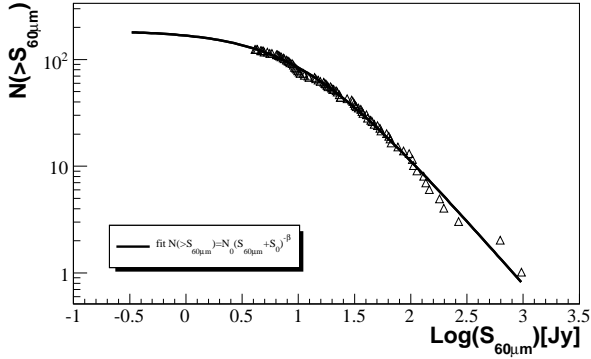}
\caption{$log(N) - log(S)$ representation of the catalog. An $S^{-1.2}$-fit matches the data nicely, with a turnover at $S_{0}=10.56\,\mbox{Jy}$.}
\label{ngroessers}
\end{figure}
Figure \ref{ngroessers} shows the logarithmic number of sources above an FIR flux density $S_{60\mu}$. The data was fitted with the following function:
\bq
N(>S) = N_0 \cdot (S+S_0)^{-\beta}
\eq
Here, $N_0$, $S_0$ and $\beta$ are the fit parameters. Using an error of $\sqrt{N}$, the parameters are determined to
\bqan
N_0   &=& 3155 \pm 1297.9\\
S_0   &=& (10.56 \pm 3.78)\,\mbox{Jy}\\
\beta &=& 1.2 \pm 0.2.
\eqan
The behavior $N(>S)\sim S^{-1.2\pm0.2}$ matches the expectation that the function should lie between $S^{-1.0}$ and $S^{-1.5}$.\\
The next section will explain the connection between starburst galaxies and gamma ray bursts which will result in a flux estimation for high energy neutrinos
from GRBs inside starburst galaxies.
\section{Gamma Ray Bursts and starbursts}
\label{grb_sb}
\begin{figure}
\includegraphics[width=\columnwidth]{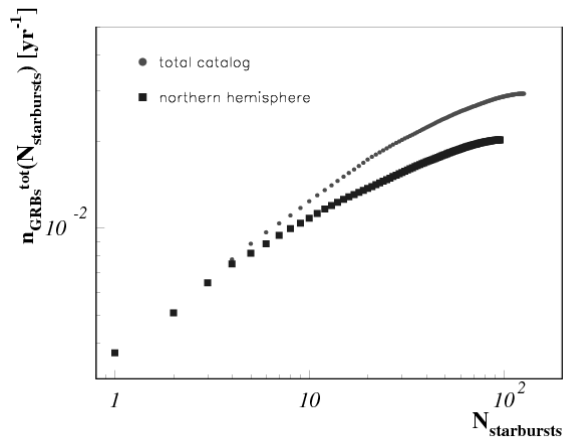}
\caption{Number of GRBs per year in the starburst catalog, including $N_{\rm{starbursts}}$ sources, starting with the strongest one. 
The total GRB rate in the sample, including all $127$ sources is $0.03\,\mbox{yr}^{-1}$, this means that a GRB could be observed every $30$ years on average. 
The total GRB rate just in the northern hemisphere is $0.02\,\mbox{yr}^{-1}$ or an occurrence every $50$ years, the data shown in squares. These source lie in IceCube's FoV.}
\label{NGRBSB}
\end{figure}

Starburst galaxies show an enhanced rate of supernova explosions due to their large star formation rate. Thus an increasing rate of
long Gamma Ray Bursts (GRBs) directly linked to SN-Ic events \cite{Mazzali2003} is expected. If long GRBs are the dominant sources of
UHECRs, the contribution from nearby objects should follow the distribution of starburst galaxies. In the following calculations, it is assumed that every
SN-Ic explosion is accompanied by a particle jet along the former star's rotation axis, i.e. by a GRB. The opening angle of the GRB jet, $\theta$ determines, how many SN-Ic 
can be observed as GRBs, see e.g. \cite{Berger,Racusin}. 
\bq
\dot{n}_{\rm{GRB}} = \epsilon \cdot \dot{n}_{\rm{SN-Ic}}.\label{ngrb}
\eq
Here, $\dot{n}_{\rm{GRB}}$ is the GRB rate in a galaxy and $\epsilon = (1-\cos(\theta))$ is the fraction of SN-Ic which can be seen as GRB.
The opening angle is expected to be less than $\sim 10^{\circ}$ for
the prompt emission. Afterglow emissions and
precursors can have a larger opening angles \cite{Morsony}. Putting the focus on prompt emission, an optimistic opening angle
of $\sim 10^{\circ}$ is used, yielding
\bq
\epsilon = 0.015.
\eq
Further, observational data show that core collapse supernovae of type Ic contribute with $11\%$ to the total supernovarate in starbursts \cite{Cappellaro}. Using equation \ref{ngrb}
the GRB rate in a starburst galaxy is directly correlated to the supernova rate $\dot{n}_{SN}$,
\bq
\dot{n}_{\rm{GRB}}=\epsilon \cdot \xi \cdot \dot{n}_{\rm{SN}}\label{ngrbsn}
\eq
with $\xi\sim 0.11$ as the fraction of heavy SN among all SN. The supernova rate is correlated with the FIR luminosity of the galaxy \cite{Mannucci},
\bq
\dot{n}_{\rm{SN}}=(2.4\pm0.1) \cdot 10^{-12}\cdot \left( \frac{L_{\rm{FIR}}}{L_{\odot}} \right)\,\mbox{yr}^{-1}.\label{nsnfir}
\eq
The FIR luminosity is expressed in terms of the solar luminosity $L_{\odot}=3.839 \cdot 10^{33} \mbox{erg\/s}$ and is given in the range of $60\,\mu\mbox{m}$ and $100\,\mu\mbox{m}$ by \cite{Xu}
\bq
L_{\rm{FIR}} = 4 \pi d_l^2 \cdot F_{\rm{FIR}}.
\eq
Here, $d_l$ is the luminosity distance of the individual source and
\bq
F_{\rm{FIR}} = 1.26 \cdot 10^{-14} \cdot \left( 2.58\cdot S_{60/\mu} + S_{100\mu} \right) \mbox{W}\,\mbox{m}^2
\eq
is the FIR flux density as defined in \cite{Helou}. $S_{60\mu}$ and $S_{100\mu}$ are the measured flux densities at $60\,\mu\mbox{m}$ and $100\,\mu\mbox{m}$, both measured in Jy.
Using equation \ref{nsnfir} to determine the supernova rate, equation \ref{ngrbsn} yields a rate of
\bq
\dot{n}_{\rm{GRB}}=3.8 \cdot 10^{-15} \cdot \left(\frac{L_{\rm{FIR}}}{L_{\odot}}\right)\cdot \left(\frac{\epsilon}{0.015}\right) \cdot \left(\frac{\xi}{0.11}\right)\,\mbox{yr}^{-1}
\eq
per starburst. For a $1\,\mbox{km}^3$ neutrino detector with a lifetime of $10$ years (like IceCube or KM3NeT) luminosities of around $3\cdot 10^{13}\cdot L_{\odot} \sim 10^{47}\,\mbox{erg\/s}$ are required for a 
single event within these $10$ years.
None of the sources in the catalog provides such a high luminosity. However, if a larger number of starburst is considered for an analysis, the total luminosity increases and so does the probability of observing a GRB.
Figure \ref{NGRBSB} shows the total GRB rate for a number of $N_{\rm{starbursts}}$ galaxies,
\bq
\dot{n}^{\rm{tot}}_{\rm{GRB}}(N_{\rm{starbursts}})=\sum_{i=1}^{N_{\rm{starbursts}}} \dot{n}_{\rm{GRB}}(i\mbox{th starburst}).
\eq
In the figure the GRB rates achieved in the single starbursts are summed up, starting with the most luminous source, adding sources in descending luminosity order. 
The points show the GRB rate summing up over all starbursts in the
sample, starting with the strongest one and adding the next strongest
sources subsequently. On total, $0.03$ GRBs per year are expected to be observable in the sample. The squares display the total GRB rate, 
summing up sources in the northern hemisphere, which corresponds to IceCube's Field of View (FoV). Here, 0.02 GRBs per year are expected. This number can be enhanced significantly
when taking weaker sources into account which were not included in the sample in order to ensure completeness, see section \ref{sample}.
\section{Enhanced neutrino flux from GRBs in starbursts}
\label{nuflux}
Due to the high atmospheric background seen by high-energy neutrino telescopes, the detection of a diffuse neutrino signal from GRBs in nearby starbursts will not be possible.
However, with a timing analysis one might be able to identify Gamma Ray Bursts in neutrinos. In such an analysis, the location of a nearby starburst can be chosen as a potential neutrino hot-spot.
By selecting a time window of the typical duration of a long GRB ($\sim 100\,\mbox{s}$) the atmospheric background can then be reduced to close to zero. 
In this context the general neutrino intensity and in particular the possibility of neutrino detection with IceCube are discussed.\\
For the first time the neutrino energy spectrum during the prompt photon emission phase in a GRB
was determined by Waxman\&Bahcall \cite{WB97,WB99} and can be expressed as 
\bq
\frac{dN_{\nu}}{dE_{\nu}}=A_{\nu}\cdot E_{\nu}^{-2} \cdot \left\{
\begin{array}{ll}
E_\nu^{-\alpha_{\nu}+2}\cdot {\epsilon_{\nu}^{b}}^{\alpha_{\nu}-\beta_{\nu}} &, E_{\nu}<\epsilon_{\nu}^{b}\\
E_\nu^{-\beta_{\nu}+2}                                                       &, \epsilon_{\nu}^{b} < E_{\nu} \le \epsilon_{\nu}^{S}\\
\epsilon_{\nu}^{S}\cdot E_\nu^{-\beta_{\nu}+1}                               &, E_{\nu}>\epsilon_{\nu}^{S}\\
\end{array}
\right.
\eq   
The spectrum includes two spectral indices,  $\alpha_{\nu}$ and $\beta_{\nu}$, two break energies, $\epsilon_{\nu}^{b}$ and $\epsilon_{\nu}^{S}$ and a
normalization factor $A_{\nu}$. For the GRBs in starbursts, these parameters were discussed in detail in \cite{CR6}. Their numerical values were determined to
\bqan
\alpha_{\nu}       &=& 1\\
\beta_{\nu}        &=& 2\\
\epsilon_{\nu}^{b} &\approx&3\cdot 10^{6}\,\mbox{GeV}\\
\epsilon_{\nu}^{S} &\approx&3\cdot 10^{7}\,\mbox{GeV}\\
A_{\nu}            &\propto& d_{l}^{-2}
\eqan
The normalization constant $A_{\nu}$ is calculated for each individual source. It depends on $d_{l}^{-2}$ as well as on the
fraction of energy transferred into electrons and the fraction of
energy transferred into charged pions. In addition, the normalization
of the neutrino spectrum scales with the luminosity of the
burst. This released energy varies from burst to burst. In addition to
this burst-to-burst fluctuation,
regular GRBs are distinguished from low-luminosity bursts. Regular, long bursts emit a total isotropic energy of
$10^{52}\,\mbox{erg}$ for a duration ($t_{90}$) of the burst of
$\approx 10\,\mbox{s}$. Low-luminosity bursts last longer and and have a lower
luminosity. Although only few low-luminosity bursts
are observed yet, they are expected to be much more frequent than
regular GRBs. For
this class, we expect an energy release of $\sim 10^{50}$~erg within
around $1000$~s. The closest burst observed so far was GRB980425, which
was found to be associated with the supernova SN1998bw \cite{galama1998}. The host
galaxy lies at a redshift of only $z=0.0085$. This burst shows a total
energy release of $\sim 10^{47}$~erg, which is an extremely
low-luminosity burst. As the luminosity distribution is not well-known
at this point, due to low statistics, we use a fixed value of
$10^{51}$~erg. An actual burst can be about one order of
magnitude more or less luminous.
Now, to estimate the neutrino flux from a standard GRB for a single starburst in the sample, dependent on the distance
of the starburst, the normalization is calculated. The other parameters are kept constant and hence the results can only serve as a rough estimate. Both, the break energies as well as
the spectral indices vary for each individual burst as described in \cite{Guetta}.
\subsection{Expected event rates in IceCube}
\label{ic3}
As shown in figure \ref{NGRBSB}, $0.02$ GRBs per year are expected to occur in the $96$ starbursts in the sample in the northern hemisphere, IceCube's FoV.
However, this rate can be enhanced if all sources in the northern hemisphere would be taken into account. For
completeness reasons only the brightest ones were considered here. This enhances the possibility to detect a GRB from a starburst in the super galactic plane within the lifetime
($10$~years) of IceCube. The prospects for KM3NeT are slightly worse, since only $0.01$ GRB per year is expected in the southern hemisphere, where KM3NeT's FoV will
be focused.

\begin{figure}
\includegraphics[width=\columnwidth]{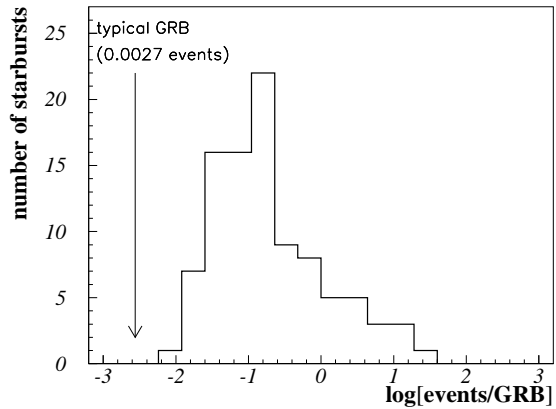}
\caption{Histogram of the number of events in IceCube from the $96$ starburst galaxies in the northern hemisphere.
Depending on the distance of the starburst, a burst would result in between $0.1$ and several $100$ events in IceCube in a small time window of $10-100$ seconds. A regular burst
at $z\sim 1-2$ gives only $0.027$ events as indicated in the figure. The main reason for the increased signal is that the
bursts would come from starbursts closer than $z=0.03$.}
\label{NEVIC3}
\end{figure}

The number of events per GRB expected in IceCube can be calculated by folding IceCube's effective area $A_{\rm{eff}}$~\cite{Montaruli}
with the GRB spectrum $dN_{\nu}$/$dE_{\nu}$
\bq
N_{\rm{events}} = \int_{E_{\rm{th}}}^{\infty} A_{\rm{eff}}(E_{\nu}) \cdot \frac{dN_{\nu}}{dE_{\nu}}\,dE_{\nu}
\eq 
Here, the weak dependence of the $A_{\rm{eff}}$ from the declination of the burst is neglected. For the threshold energy
$E_{\rm{th}}=100\,\mbox{GeV}$ is used. This is the general detection threshold of IceCube \cite{Ahrens}. Since events
can be selected in a small time window with the typical duration of a long GRB, $10-100\,\mbox{s}$, the atmospheric background
can be reduced close to zero. Figure \ref{NEVIC3} shows the histogram of the numbers of events expected in IceCube from an average
GRB with an isotropic energy of $E_{\gamma}^{\rm{iso}}=10^{51}\,\mbox{erg}$ from the $96$ starbursts in the sample.
These numbers range from $0.1$ to about $300$ events per burst, depending on the distance of its host galaxy. These numbers lie between $1$ and $5$ 
orders of magnitude above the numbers of events for GRBs typically observed by satellite experiments like Swift, BATSE and Fermi. If such GRB occurs
in our sample of starburst galaxies it is a unique opportunity to study the hadronic component of GRBs.
\section{Conclusions and Outlook}
A sample of $127$ nearby starburst galaxies has been presented, where
we selected sources with $z<0.03$ and radio (FIR) luminosities above
$S_{60\mu}>4\,\mbox{Jy}$ and $S_{1.4\,\rm{GHz}}>20\,\mbox{mJy}$.
\newpage
The expected event rates for high energy neutrinos in the IceCube detector 
have been calculated to reach up to more than 100 events per
year. 
If a GRB occurs in one of the $127$ starbursts,
for detectors with a wide field of view the detection probability is
best. IceCube should be able to see the strongest events ($\sim 100$
neutrinos) immediatly,
while the weaker events ($\sim 5-10$ neutrinos) can be found in a
specific analysis for the selected starbursts. 
Since most of the sources are located in the northern hemisphere, detectors like
IceCube, HAWC and Auger north are optimal for such a study. 
A detection of neutrinos from a GRB would enable detailed studies
of hadronic emission processes of GRBs.
%
%
\section*{Acknowledgments}
We would like to thank Jay Gallagher, Francis Halzen and Kotha Murase for helpful
discussions. JD, JKB and WR are supported by the IceCube grants
BMBF (05~CI5PE1/0) and (05~A08PE1). J.K.~Becker and J.~Dreyer acknowledge 
the support from the 'Research Department of Plasmas with complex interactions'. 


%
%

\end{document}